# High speed optical processing via four-wave-mixing in a silicon waveguide

F. Li[1], M. Pelusi[1], D-X. Xu[2], A. Densmore[2], R. Ma[2], S. Janz[2], and D.J. Moss[1]

[1] *Institute of Photonics and Optical Sciences (IPOS) and the Centre for Ultra-high Bandwidth for Optical Devices (CUDOS), School of Physics, University of Sydney, New South Wales 2006 Australia*
[2] *Institute for Microstructural Sciences, National Research Council, Ottawa, Ontario, Canada K1A-0R6*
dmoss@swin.edu.au

**Abstract** We demonstrate all-optical demultiplexing at 160Gb/s in the C-band, based on four-wave mixing (FWM) in a silicon nanowire. We achieve error-free operation with a penalty of ~ 3.9dB at $10^{-9}$ bit error rate.

## 1. Introduction

All-optical nonlinear signal processing is seen [1,2] to be critical for future telecommunication networks to address the growing demand for network flexibility, low cost, energy consumption and bandwidth as they evolve to well beyond the capability of electronic device speeds towards 640Gb/s [3], Terabit Ethernet [4], and beyond. Nanophotonics, particularly in highly nonlinear materials, has been a key approach to reducing operating power requirements of all-optical devices, by increasing the nonlinear parameter, $\gamma = \omega\, n_2 / c\, A_{eff}$, where $A_{eff}$ is the waveguide effective area. Both silicon and chalcogenide glass (ChG) waveguides have achieved extraordinarily high $\gamma$ 's ranging from 1.2 $W^{-1} m^{-1}$ in chalcogenide fiber [5], 15 $W^{-1} m^{-1}$ in ChG waveguides [6,7] to 95$W^{-1} m^{-1}$ in ChG tapered fiber nanowires[8], to 300$W^{-1} m^{-1}$ in Si nanowires [9]. This has led to impressive demonstrations of all-optical demultiplexing at 160Gb/s in ChG fibre nanowire tapers [10] to 640Gb/s demultiplexing in ChG waveguides [3] as well as optical spectrum measurement to beyond 1THz [11]. Many demonstrations of all-optical signal processing have also been reported in silicon devices [9, 12-24] from wavelength conversion to switching, demultiplexing and regeneration, parametric gain and many others. However, the current record speed is 40Gb/s [12, 13], even though in principle these devices are intrinsically capable of operating at much higher speeds. In particular, error-free transmission in a silicon nanowire of multiple wavelength division multiplexed channels with an aggregate data transfer of 1.28Tb/s has been reported [16]. Further, switching in ring resonators based on free-carrier injection in a cascaded ring optical waveguide (CROW) device has also been reported [17], where a switching speed of ~ 2nS was achieved on individual channels running at a bit rate of 40Gb/s.

In this paper, we report record high bit rate operation for all-optical signal processing in a silicon device. We demonstrate time division switching, or de-multiplexing, of a 160Gb/s (33% Return-to-Zero) data stream ($2^{31}$-1 pseudorandom bit sequence (PRBS)) in the C-band, down to 10Gb/s, via four-wave mixing (FWM) in a silicon nanowire. We perform system penalty measurements and achieve error free operation with a system penalty of ~ 3.9dB at a bit error rate (BER) of $10^{-9}$. Our results show that free-carrier dynamics, whilst they do adversely influence device performance, do not pose a barrier to device operation at extremely high bit rates, as the timescale for carrier recombination is much longer than the inter-bit period.

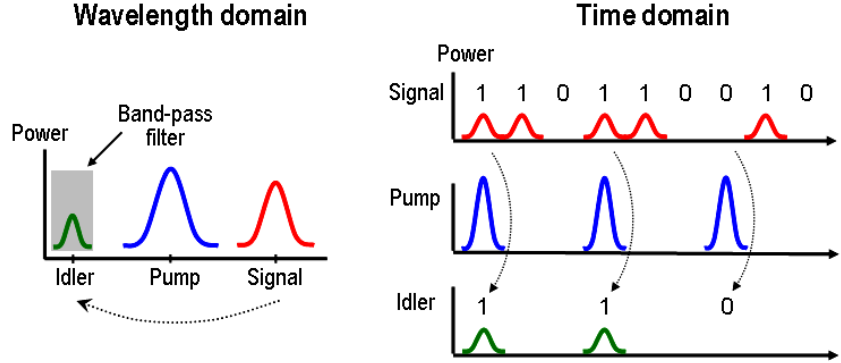

Figure 1 shows the principle of FWM-based demultiplexing [10,20]. A high bit rate signal (frequency $f_s$) is co-propagated with a sub-harmonic pump pulse train (frequency $f_p$ ). FWM between coinciding signal and pump pulses generates idler pulses at a carrier frequency $f_i = 2f_p - f_s$ that is filtered and directed to a receiver (or re-routed) without electronic conversion or regeneration.

## 2. Experiment

Our device is a 1.1cm long silicon-on-inslulator (SOI) "nanowire" with dimensions of 450nm wide x 260nm thick (Figure 2), fabricated by electron-beam lithography and reactive ion etching on an SOI wafer with a 2 µm buried oxide layer. The waveguides were coated with 2 µm thick SU8 polymer layer which has a refractive index of 1.575 at $\lambda = 1550$ nm . To improve fiber to waveguide coupling efficiency, a nominal 450 nm wide waveguide was linearly tapered to 150 nm at the facet for adiabatic expansion of the waveguide mode. This inverse taper design was optimized for TM operation and has a high loss for TE due to leakage into the silicon substrate [25]. The measured propagation loss was ~ 4dB/cm for both TE and TM polarizations and coupling losses were 4.5dB/facet for TM and 7.5dB/facet for TE polarizations (coupling via lensed fiber tapers using nanopositioning stages) resulting in a total insertion loss of 19dB (TE pol.) and 13dB (TM pol.). The total dispersion (waveguide plus material) is anomalous and roughly constant at ~  + 500ps/nm/km over the C-band for TE, and for TM it is normal and varies between approximately ~ -8,000 ps/nm/km and ~ -16,000 ps/nm/km.  We therefore used TE polarization for these experiments despite the coupling loss being higher.

The experimental setup for demonstrating demultiplexing from 160Gb/s to 10Gb/s is shown in Fig. 2. The 160 Gb/s signal was generated from a 40 GHz mode-locked fiber laser emitting 1.4 ps, 1.8 nm bandwidth pulses at $\lambda = 1565$ nm. A Mach-Zehnder electro-optic modulator encoded data on the pulses at 40 Gb/s with a $2^{31}-1$ PRBS, and a two-stage multiplexer (MUX) of $2^7-1$ bit delay-length interleaved the signal up to 160 Gb/s (pulse width 1.9 ps, 30 % duty cycle). The pump pulse train was sourced from a 10 GHz modelocked fiber laser (2.3 ps long, 1.1 nm bandwidth at 1546 nm) with an incident average power = 44mW, corresponding to 11mW average power, or 500mW peak power, inside the waveguide). The clock was synchronized to the 160 Gb/s signal (using the same 40 GHz RF clock), pre-scaled to 10 GHz (in place of using a 10 GHz clock recovery circuit). The (TE polarized) signal and pump were combined with a 50:50 coupler and amplified by an erbium-doped fiber amplifier (EDFA) before launching into the waveguide with polarizations aligned via polarization controllers (PC). An optical delay line (ΔT) aligned the pump pulses with the channel of the 160 Gb/s signal to be demultiplexed. The average signal power was 4 mW (in the waveguide, or ~16 mW peak incident) for a combined (pump+signal) average launch power of 60 mW. We performed system penalty measurements [6] in order to evaluate the transmission bit error ratio, or rate (BER).

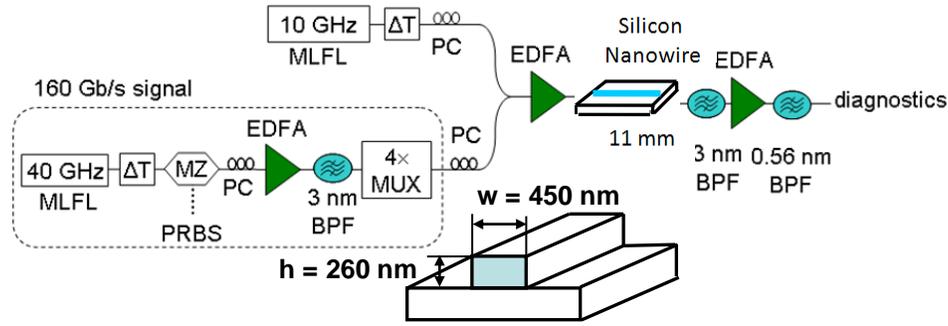

Fig.2 Experimental setup for demultiplexing from 160Gb/s to 10Gb/s (33% RZ). Note the SU8 over-cladding layer for the nanowire is not shown.

## 3. Results and discussion

The optical spectra at the waveguide output (Figure 3) shows the pump and signal spectra ( ~ 20 nm wavelength separation) as well as the idler spectrum at 1528nm generated by FWM. When the reduction in pulse repetition rate from 160Gb/s to 10Gb/s (at the idler) is accounted for, we obtain an on/off (neglecting losses) conversion efficiency of ~ -9dB, which compares favorably to other reports [12, 13]. Figure 4 shows the eye diagrams of the input 160Gb/s data stream along with the 10Gb/s pump beam at 1546nm as well as the output optimized eye diagram of the demultiplexed 10 Gb/s idler, all extracted by tunable bandpass optical filters. All traces were measured with a 65GHz detector and sampling oscilloscope. These results highlight the effective demultiplexing operation that has been achieved. Finally, we performed system BER measurements on the device. Figure 5 shows the system penalty measurements indicating a penalty, relative to back-to-back operation, of about 3.9dB at $10^{-9}$ BER. We anticipate a significant reduction in both the system penalty and operating pump power by reducing coupling and propagation losses as well as optimizing the waveguide dispersion closer to the ideal for FWM, namely, small but anomalous [14]. Coupling loss can be significantly reduced with full tapered couplers (down to a few 10's of nm, to expand the mode further), rather than the partial tapers used here.

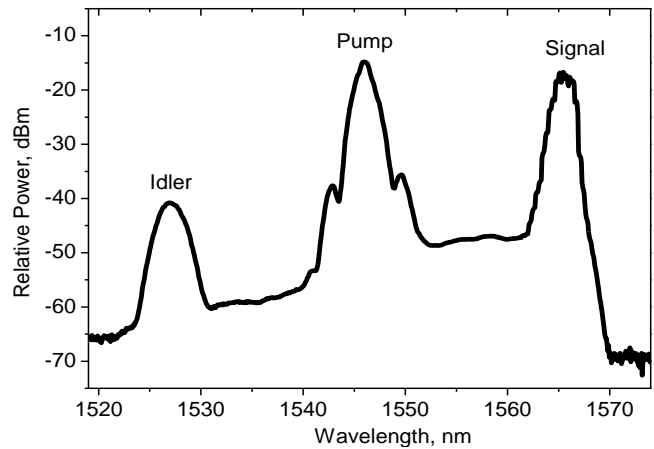

Fig.3 Output spectrum measured on an OSA, collected at the DROP port.

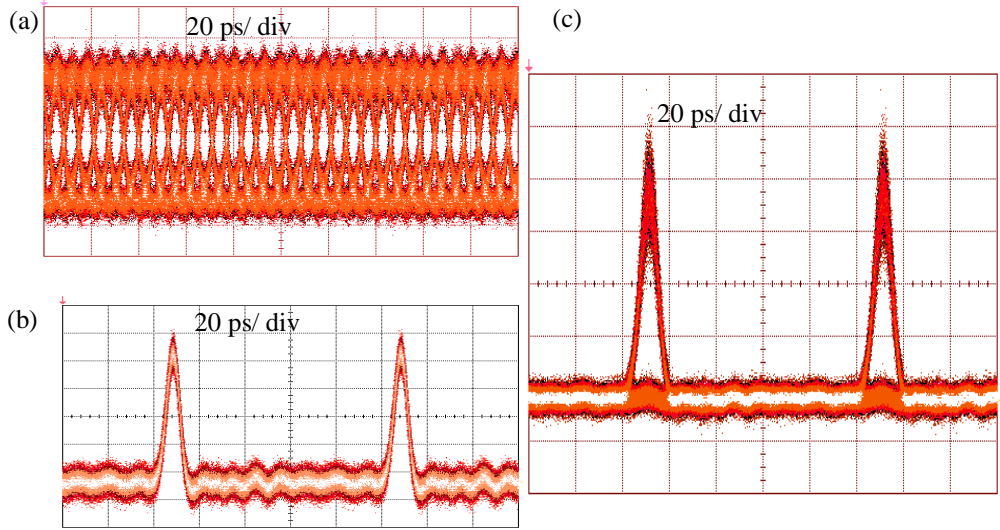

Fig.4 Eye diagrams corresponding to (a) the 160Gb/s input signal at λ = 1565 nm, (b) 10Gb/s pump beam at λ = 1546nm and (c) 10Gb/s output idler at λ = 1528nm .

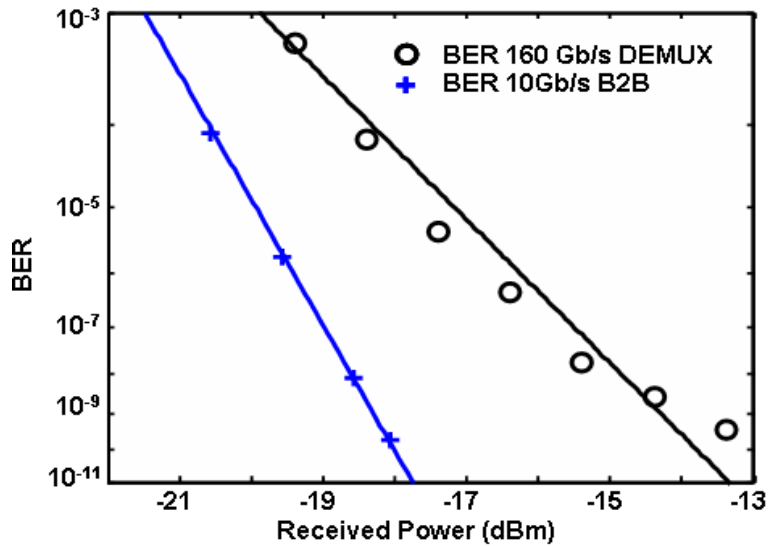

Fig.5 BER measurement for the signal (blue line) and the idler (black line) generated by a 4mW signal and a 11mW pump: the power penalty of the two curves is 3.9dB.

The error-free operation of our device clearly demonstrates that free-carriers are not a barrier to effective device operation. This is not unexpected since the source of free carriers in our case is the pump, which is a 10Gb/s continuous clock, with no pattern dependent information. Also, these effects will be much smaller than at 40Gb/s, where error-free wavelength conversion via FWM in silicon nanowires has been demonstrated [12,13]. We note that at 160Gb/s the bit period of 8ps is ~ 1/100 of the typical carrier lifetime, and the probability of obtaining 100 consecutive 1's (or 0's) is negligible ( $< 10^{-30}$). Notwithstanding this, it is clear that reducing free carrier effects will increase the device efficiency and reduce operating power requirements. Detailed modeling on this will be presented in a future work.

## 4. Conclusions

We report record high bit rate operation for all-optical signal processing in a silicon device. We demonstrate time division de-multiplexing of a 160Gb/s (33% Return-to-Zero) PRBS data stream in the C-band, down to 10Gb/s, via FWM in a silicon nanowire. We perform bit error ratio measurements and achieve error free operation with a penalty of 3.9dB at $10^{-9}$ BER.

## Acknowledgements

We acknowledge financial support of the Australian Research Council (ARC).